# Ruthenium-Europium configuration in the Eu$_2$Ru$_2$O$_7$ pyrochlore

S. Muñoz Pérez,[1] R. Cobas,[1] J. M. Cadogan,[1] J. Albino Aguiar,[2] S. V. Streltsov,[3,4] and X. Obradors[5]

[1]*School of Physical, Environmental and Mathematical Sciences, UNSW Canberra at the Australian Defence Force Academy, Canberra, ACT 2610, Australia*
[2]*Departmento de Fisica, Universidade Federal de Pernambuco, Recife 50670-901, Brazil*
[3]*Institute of Metal Physics, Ekaterinburg 620219, Russia*
[4]*Ural Federal University, Ekaterinburg 620002, Russia*
[5]*Institut de Ciència dels Materials de Barcelona, CSIC, Campus de la UAB, 08193 Bellaterra, Spain*

The magnetic and electronic properties of Eu$_2$Ru$_2$O$_7$ are discussed in terms of the local ruthenium and europium coordination, electronic band structure calculations and molecular orbital energy levels. A preliminary electronic structure was calculated within the LDA and LSDA+U approximations. The molecular orbital energy level diagrams have been used to interpret the Eu-Ru ligand spectrum and the ensuing magnetic properties. The orbital hybridizations and bonds are discussed.

## I. INTRODUCTION

Ruthenium compounds with the pyrochlore structure A$_2$Ru$_2$O$_6$O′ (where A = rare earth) exhibit a lambda-like specific heat jump at a temperature T$_1$ (between 60 K and 170 K) due to a quasi-long-range antiferromagnetic (AFM) ordering of the Ru$^{4+}$ sublattice.[1] This magnetic ordering of the Ru sublattice creates a molecular field at the A site which polarizes the rare earth (RE) magnetic moments and is sufficient to induce magnetic order in all the magnetic RE ions such as Tb, Yb, Er, Gd and Ho.[2-7] In Eu$_2$Ru$_2$O$_7$ the Ru sublattice orders at T$_1$, as expected, but a second prominent magnetic transition occurs at a lower temperature despite the fact that Eu$^{3+}$ is presumed to be in its non-magnetic $^7F_0$ ground state.[8,9] Due to the usual non-magnetic nature of the Eu$^{3+}$ ions, one would expect an intrinsic behaviour similar to that observed in the Y and Lu pyrochlores where no magnetic transitions occur at low temperatures.[10] This anomalous second magnetic transition has been systematically reported in the literature.[7-8] However, due to difficulties in performing neutron scattering experiments on Eu-based compounds, the precise nature of the observed ordering remains unknown.

## II. EXPERIMENTAL RESULTS

The Eu$_2$Ru$_2$O$_7$ powder sample was prepared by solid state reaction as described in Ref. 9. Structural characterization was done by synchrotron X-ray powder diffraction (SXRD) using λ= 0.39468(1) Å.

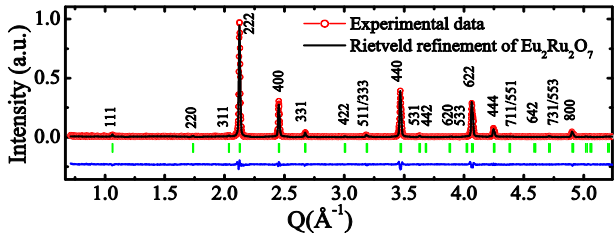

FIG 1. SXRD pattern of Eu$_2$Ru$_2$O$_7$ obtained at 300 K.

Rietveld analysis of the SXRD pattern obtained at 300 K, using the FullProf suite, shows single-phase Eu$_2$Ru$_2$O$_7$ with the $Fd\bar{3}m$ space group (see Fig. 1a). No crystallographic transition was observed at low temperatures.

The dc-susceptibility and transport measurements were made in a Superconducting Quantum Interference Device (SQUID) magnetometer and a Physical Properties Measurements System (PPMS) respectively. The Ru sublattice orders at T$_1$ ~ 118 K and a second major divergence between the field-cooled (FC) and zero-field cooled (ZFC) susceptibilities occurs below T$_2$ ~ 23 K as shown in Fig. 2a.

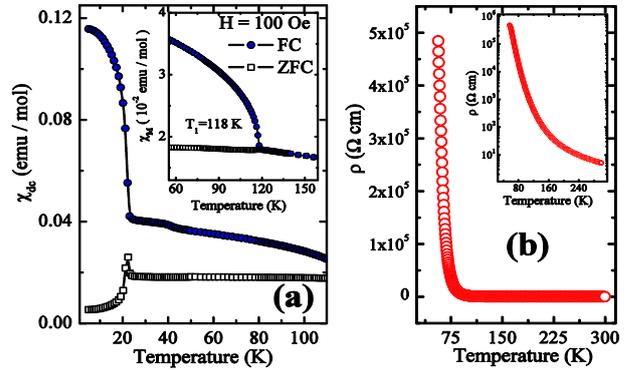

FIG 2. (a) Magnetic $\chi_{dc}$ of Eu$_2$Ru$_2$O$_7$ measured under ZFC and FC conditions. (b) Temperature dependence of the resistivity of Eu$_2$Ru$_2$O$_7$.

The temperature dependence of the electrical resistivity indicates a semiconductor to insulator transition with a dramatic increase in resistivity below 75 K (see Fig. 2b).

## III. THEORETICAL RESULTS AND DISCUSSION

The magnetic and electronic properties of Eu-pyrochlore compounds are special amongst all the RE because of the particular 4$f^6$ energy level structure of the Eu$^{3+}$ ions.

In the Eu$_2$Ti$_2$O$_7$ pyrochlore it has been shown that the europium ions experience a nearest-neighbour exchange interaction with a weak dipolar interaction below 4 K.[11-12] More interestingly, luminescence studies of Eu$_2$Ti$_2$O$_7$ and Eu$_2$Sn$_2$O$_7$ reveal that after excitation into the $^5D_1$ level, the emission lines below 20 K are in the $^5D_0\rightarrow{}^7F_0$, $^5D_0\rightarrow{}^7F_1$, and $^5D_0\rightarrow{}^7F_2$ spectral region. At low temperatures, the emission peaks are dominated by trap emission due to direct energy transfer from intrinsic $Eu^{3+}$ ions to the extrinsic europium ions. The nature of the extrinsic europium has been ascribed to a considerable degree of anion disorder, covalency/ligand polarizability and vacancies.[13-15]

In the pyrochlore structure, the $Fd\bar{3}m$ symmetry fixes the positions of all ions except for the oxygen in the 48f site ($O_{48f}$).[7,16,17] The 16$d$ site is coordinated by six $O_{48f}$ ions and two O′ ions which might suggest that cubic symmetry would be a good approximation. However, the site is strongly distorted from 8-fold cubic. The $x_{48f}$ parameter determines the degree of distortion of the $REO_8$ cube and $RuO_6$ octahedron as a result of a compression or expansion through the $C_3$ symmetry axes.

In $Eu_2Ti_2O_7$, $Eu_2Ru_2O_7$ and $Eu_2Sn_2O_7$ the $Eu^{3+}$ ions occupy the 16d lattice site with a point symmetry $\bar{3}m$ ($D_{3d}$). However, local disorder arises when some of the O' ions occupy the 8b positions ($O_{8b}$), leaving the site vacant. This creates perturbed sites having different energies for the adjacent europium ions.[13-14] Consequently, the $D_{3d}$ symmetry of the perturbed ions is lowered to $C_{3v}$ and based on the Wybourne-Downer mechanism the presence of a small crystal field (CF) term in the Hamiltonian could induce the transitions $^5D_0 \rightarrow {}^7F_0$ and $^5D_0 \rightarrow {}^7F_2$ due to the CF induced $J$-mixing effect.[16-20] On the other hand, according to the Judd–Ofelt theory, the $^5D_0 \rightarrow {}^7F_1$ electric dipole transitions are forbidden, but this is not the case for a magnetic dipole transition.[21,22] Therefore, the observations, at low temperature, of emission lines in the spectral region $^5D_0 \rightarrow {}^7F_1$ and energy transfer mechanism suggest that the interaction between the europium ions is exchange-based in character.

Since the structural parameters of $Eu_2Ru_2O_7$ are intermediate between those of the isomorphous $Eu_2Ti_2O_7$ and $Eu_2Sn_2O_7$, the observed decrease of the FC $1/\chi$ below 23 K, might be associated with such an exchange mechanism present in $Eu_2Ti_2O_7$ and $Eu_2Sn_2O_7$. In addition, the $Ru^{4+}$ ions create an exchange field acting on the $Eu^{3+}$ ions which significantly enhances the europium exchange interactions, as demonstrated by the dramatic increase in ordering temperature for $Eu_2Ru_2O_7$ (23 K) compared to $Eu_2Ti_2O_7$ (4.9 K).[8,9,11]

Figure 3 shows a view along the z-axis of the 3D kagomé lattice for the Eu, Ru and $O_{48f}$ atoms. Molecular orbitals have been simulated for the $^7F_{1,2}$ ($m_l = -2$ and $m_l = -3$) states corresponding to the extrinsic europium. These orbitals were arranged directed towards the ruthenium polyhedron through the $C_3$ axes. This likely spin-orbital correlation, between the localized spins of the ruthenium atoms and the $^7F_j$ electrons in $Eu_2Ru_2O_7$, undoubtedly merits comprehensive studies.

*A. Electronic Calculations.*

A preliminary $Eu_2Ru_2O_7$ electronic structure was calculated within the LDA and LSDA+U approximations[23] using the Stuttgart TB-LMTO-ASA code.[24] The muffin-tin sphere radii were taken to be 2.87 a.u., 2.36 a.u., and 2.05 a.u. for Eu, Ru, and O, respectively. In our LMTO calculations we used the von Barth-Hedin exchange correlation potential.[25]

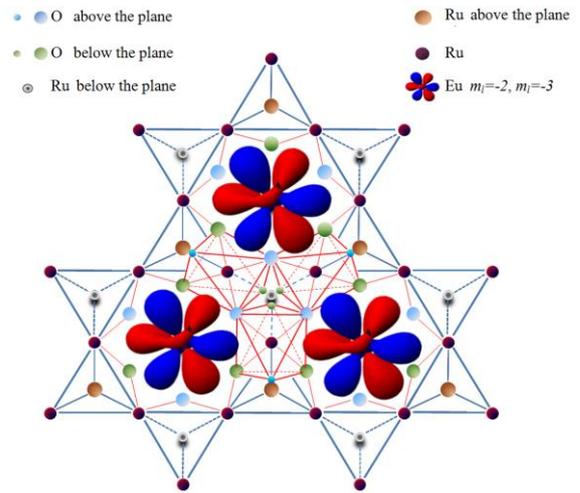

FIG. 3. View along the z-axis of the 3D kagomé lattice for $Eu_2Ru_2O_7$.

The Eu(6s,6p,5d,4f), O(2s,2p,3d) and Ru(5s,5p,4d) orbitals were included in the basis set. The Brillouin-zone (BZ) integration in the course of self-consistency iterations was performed over a mesh of 64 k-points. Parameters of the on-site Coulomb repulsion (U) and Hund's rule coupling ($J_H$) for Ru were calculated in Ref. 26,27. and taken to be the same (U=3.0 eV and $J_H$ =0.7 eV). For the Eu-4$f$ shell they were chosen to be U=6.8 eV and 0.7 eV. The spin-orbit interaction was not taken into account.

The calculations were performed for the crystal structure corresponding to T=5K. In spite of the absence of detailed information about the exact magnetic structure of this compound we assumed that half of the Ru ions in the unit cell has one direction of spin, with the second half - opposing. As a result, our choice of AFM structure corresponds to the situation where four out of six neighboring Ru have opposite spins. The europium ions that were chosen to be non-magnetic in the beginning of the calculation according to the usual Hund's rule did not magnetize later, whereas the local magnetic moment on the Ru ions was found to be 1.53 $\mu_B$.

An analysis of the occupation matrix shows that Ru has an intermediate spin configuration: $t_{2g}^4 e_g^0$. The single electron in the spin-minority band localizes on a nearly trigonal $a_{1g}$ orbital. The origin of the stabilization of this intermediate spin-state configuration lies in a strong crystal-field splitting $\Delta_{te}$ between the $t_{2g}$ and $e_g$ shells. From the nonmagnetic LSDA+U calculation, where U was applied only to the Eu-4$f$ states (to shift them away from the Fermi level and prevent charge overlapping), we estimated the crystal field splitting (as the difference between the corresponding centers of the bands[28]) to be $\Delta_{te}$ = 3.04 eV. For the $d^4$ configuration a high spin-state may be realized only in the case when $3J_H > \Delta_{te}$. This condition is not fulfilled for $Eu_2Ru_2O_7$. This explains the stabilization of the intermediate spin-state in the full LSDA+U calculation.

Total and partial densities of states (DOS) obtained for $Eu_2Ru_2O_7$ within the LSDA+U calculation are presented in Fig. 4. One may see that this compound is an insulator with a band gap ξ=0.18 eV (in the normal LDA, without accounting for the strong on-site Coulomb correlations, $Eu_2Ru_2O_7$ was found to be a metal). Both the top of the valence band and the bottom of the conduction band are formed by Ru-4$d$ states, showing that this compound should be considered as a Mott-Hubbard insulator in the Zaanen-Sawatzky-Allen classification scheme.[29] The O-2p states are located from -8 eV to -2 eV.

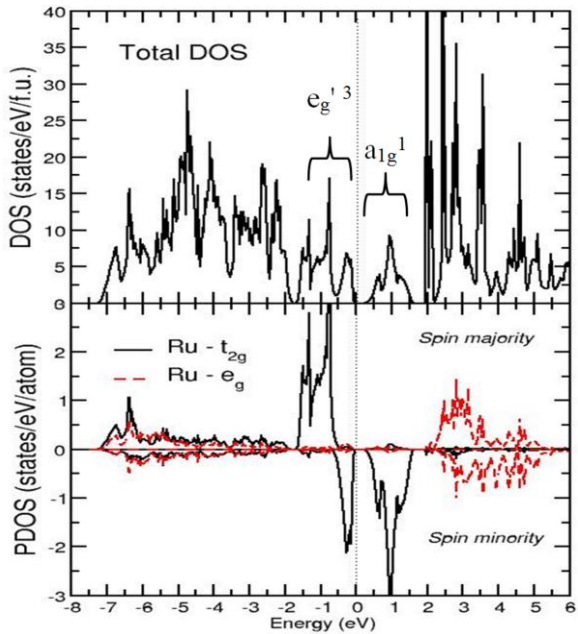

FIG.4 Total (upper panel) and partial (bottom panel) DOS for $Eu_2Ru_2O_7$ obtained within the LSDA+U calculations. Total DOS is summed over spins, while positive (negative) values of the partial DOS correspond to spin majority (minority) states. The Fermi energy is taken as zero.

In order to show that exchange field due to the Ru ions in $Eu_2Ru_2O_7$ is strong enough we calculated the exchange constants for the Heisenberg model (written as $\sum_{ij} J S_i S_j$; sum runs twice over each pair) finding the energy variation of the system under small spin rotation.[30] The exchange constant between nearest Ru neighbors is AFM and of order of 130 K.

The electronic properties calculated for $Eu_2Ru_2O_7$ could be improved by choosing different U and J values. For instance, different U values may change the occupation of the 4f orbitals.

*B. Molecular Orbital Energy Diagram.*

From the viewpoint of Molecular Orbital (MO) theory, forming a metal complex involves the reaction between a ligand (Lewis base) and a metal or metal ion (Lewis acid) to establish coordinate covalent bonds between them.[31] Our model for $Eu_2Ru_2O_7$ uses the hybridizations of the *p* and *d* orbitals to describe the magnetic and electronic properties.

The Ru-O bonds for a ruthenium atom 6-coordinated with symmetry $D_{3d}$ are achieved through hybridization, using the linear combination of three 4d ($dxy$, $dx^2-y^2$ and $dz^2$) and three 5p ($px, py, pz$) atomic orbitals to form six energy equivalents $d^3p^3$ hybrid orbitals directed towards the vertices of a trigonal antiprismatic polyhedron. Using the magnetic criterion from our electronic calculations where ruthenium has a $t_{2g}^4 e_g^0$ intermediate spin configuration, one electron must be promoted from the 4d orbitals towards the 5p as shown in Fig. 5. These hybrid orbitals have the correct symmetry and participate in the σ and π type bonds with the $O_{48f}$ to form the valence band (bonding state). The orbitals $dxz$, $dyz$ and $5s^1$ form the conduction band (nonbonding state). This configuration allows the formation of a valence band with four σ bonds (each bond consists of two electrons, one from the metal and one from the $O_{48f}$) and two π bonds. The theory also shows that electron pairs involved in a π bond are provided by the $sp^3$ hybrid oxygen orbitals, as shown in Fig 5b.

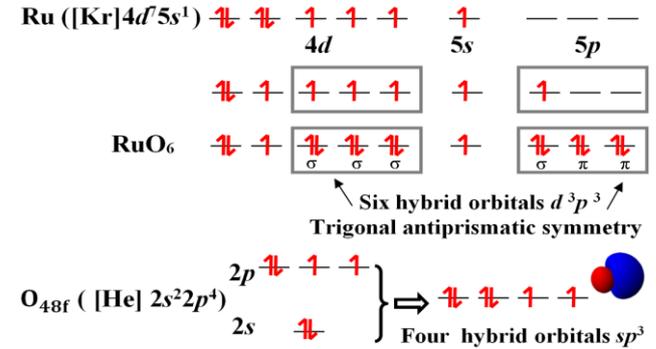

FIG. 5 Ruthenium and oxygen hybrid orbitals.

Figure 6 shows a qualitative energy level diagram for $Eu_2Ru_2O_7$. The $RuO_6$ electronic configuration considering the σ, π bonds and the $D_{3d}$ symmetric can be written as $a_{1g}^2 t_{1u}^6 e_g^4 t_{2g} (e'_g{}^3 a_{1g}^1)$ or simply $t_{2g}^4$. In agreement with the electronic calculation the energy levels below and above the Fermi level ($E_F$) have mainly $t_{2g}$ character.

The ionic/covalent bonds $O_{48f}$ - $Eu^{3+}$ - $O_{8b}$ and the excited state $^7F_1$ for the extrinsic europium ions are located at the bottom and top of the central $t_{2g}^4$ region respectively.

Our transport data obtained in the intermediate and high field regimes follow a Mott Variable Range Hopping (VRH) or Efros–Shklovskii mechanism.[32] Since the $e'_g{}^3$ electrons are highly localized, a soft Coulomb gap $\Delta_c$ of around ~ 0.29 eV can be interpreted as an increase in the density of states close to and above the $E_F$ due to the excited states from the extrinsic $Eu^{3+}$.

The d-band carriers play the major role in the electrical conductivity, however, they cannot explain the magnetic transition at $T_2$ nor the sudden increase in the resistivity at low temperature. We also observed a negative magnetoresistance above the spin-freezing temperature $T_1$ and a positive magnetoresistance below this temperature, indicative of a high degree of spin polarization of the conduction electrons at low temperatures.[9]

The interplay of such orbital degrees of freedom and spins in the pyrochlores has been known to accommodate a rich variety of phenomena. In $Eu_{2-x}Ca_xRu_2O_7$ and $Eu_2Ru_{2-x}Re_xO_7$ a delicate balance between the localized and delocalized states evolves towards a metal-like behaviour.[9,32] On the other hand, $Tl_2Mn_2O_7$ and $Tl_2Ru_2O_7$ show colossal magnetoresistance and spin gap formation, respectively.[26,33]

## IV. CONCLUSION

In summary, our results suggest that the Eu-ions play an important role in determining the electrical and magnetic properties of the $Eu_2Ru_2O_7$ pyrochlore observed at low temperatures. The calculated magnetic moment of the ruthenium ions is 1.53 $\mu_B$. The ruthenium 4d electrons promote or enhance the Eu-Eu exchange interaction, though the exact mechanism for the Eu-Ru coupling is not clear. Meanwhile, the presence of a Coulomb gap $\Delta_c$ might result in serious disruption of the electrical properties, significantly shifting the $\rho(T)$ curve from the Mott-Hubbard tendency. Therefore, a detailed experimental description of the $Eu^{3+}$ $^7F_1$ excited states will be crucial to obtain a more accurate explanation of the observed behaviour at low temperature.

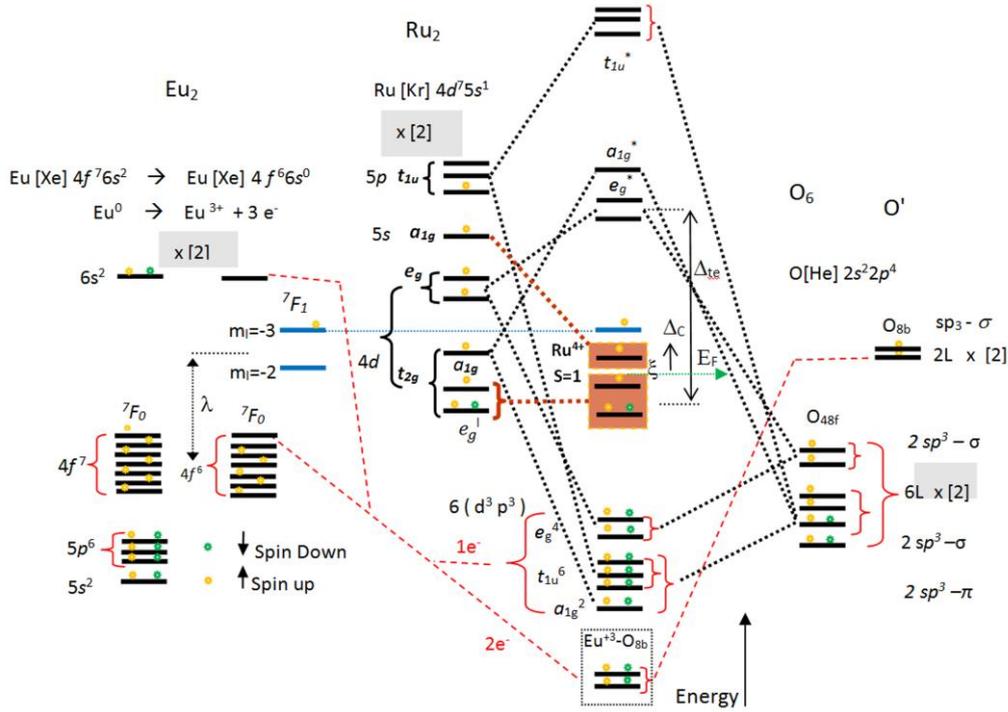

FIG 6. Schematic one-electron MO energy band diagram for $Eu_2Ru_2O_7$. The valence and conduction bands along with the electron occupancy (including spin) are indicated. Levels enclosed by parentheses are close in energy, the europium $5d$ orbital's and antibonding levels have been omitted to simplify this diagram.

**ACKNOWLEDGEMENTS**

We acknowledge financial support from the Alban Program, MICIN (CONSOLIDER NANOSELECT, CSD2007, MAT2005-02047, MAT2006-26543-E), UNSW Australia and NSERC. The analysis of the band structure, calculation of crystal field splitting and exchange correlation parameters were performed by S. Streltsov and supported by the grant of the Russian Scientific Foundation (project no. 14-22-00004).